\documentclass[aps,pre,twocolumn,groupedaddress,showpacs]{revtex4-1}
\usepackage{amsmath}
\usepackage{amsfonts}
\usepackage{amssymb}
\usepackage{blkarray}
\usepackage{graphicx}
\usepackage{xcolor}
\usepackage[hidelinks]{hyperref}
\usepackage{printlen}
\definecolor{gray-yellow}{gray}{0.886}
\definecolor{gray-cyan}{gray}{0.701}
\definecolor{gray-red}{gray}{0.299}
\definecolor{gray-green}{gray}{0.587}
\definecolor{gray-blue}{gray}{0.114}
\newlength\imageheight
\newlength\imagewidth

\begin{document}


\title{Evolutionary Fields Can Explain Patterns of High Dimensional Complexity in Ecology}


\author{James Wilsenach}
\email[]{s1666320@sms.ed.ac.uk}
\affiliation{School of Informatics, Forrest Hill, University of Edinburgh, 5 Forest Road, EH1 2QL, Edinburgh, United Kingdom}

\author{Pietro Landi}
\email[]{landi@sun.ac.za}
\author{Cang Hui}
\altaffiliation{African Institute for Mathematical Sciences, Muizenberg 7945, South Africa}
\email[]{chui@sun.ac.za}
\affiliation{Centre for Invasion Biology, Department of Mathematical Sciences, Stellenbosch University, Private Bag X1, 7602, Matieland, South Africa}

\date{\today}

\begin{abstract}
One of the properties that make ecological systems so unique is the range of complex behavioural patterns that can be exhibited by even the simplest communities with only a few species. Much of this complexity is commonly attributed to stochastic factors which have very high-degrees of freedom. Orthodox study of the evolution of these simple networks has generally been limited in its ability to explain complexity, since it restricts evolutionary adaptation to an inertia-free process with few degrees of freedom in which only gradual, moderately complex behaviours are possible. We propose a model inspired by particle mediated field phenomena in classical physics in combination with fundamental concepts in adaptation, that suggests that small but high-dimensional chaotic dynamics near to the adaptive trait optimum could help explain complex properties shared by most ecological datasets, such as aperiodicity and pink, fractal noise spectra. By examining a simple predator-prey model and appealing to real ecological data, we show that this type of complexity could be easily confused for or confounded by stochasticity, especially when spurred on or amplified by stochastic factors that share variational and spectral properties with the underlying dynamics.
\end{abstract}


\maketitle

\section{Introduction}
Complexity in ecological data is characterized by long and short-term variations in behaviour across a wide range of time-scales, from generations to speciations, which are often difficult to predict. These erratic oscillations are commonly attributed to a combination of density-dependent, demographic and environmental factors, including variation caused by human intervention \cite{bjornstad2001noisy}. However, high-dimensional deterministic effects can be difficult to distinguish from high or infinite dimensional stochasticity, especially when data sets are relatively small (as is common in ecology). These patterns of variation have characteristic spectral compositions \cite{pimm1988variability} and are often fractal in nature \cite{osborne1989finite}. Field-based models of systems with many constituent particles have been used to understand unpredictable and fractal systems found in physics \cite{hietarinta1993chaos,goldfain2006complexity,nottale1996scale}, and were central to the development of complex systems research \cite{gutzwiller1998moon}. We ask whether a field, mediated by interacting individuals in evolving populations, could adequately describe some of the properties of ecological systems seen in nature by qualitative analysis of the field-based system as a whole and at the population level. 

Dercole et al. \cite{dercole2010chaotic,derichaos} were the first to demonstrate a minimal adaptive ecological network, comprising three co-evolving species, prey, predator and super-predator, in which red queen dynamical chaos in the co-evolution of traits leads to an increase in complex behaviour at the population level \cite{dieckmann1995,Dercole_et_al_06_PRSB}. However, slower, first order evolutionary dynamics constrain the complexity, period and magnitude of such chaotic oscillations. This results in part from fundamental properties of the so-called canonical equation in adaptive dynamics (AD) which admit only first order solutions in trait space \cite{dieckmann1996dynamical,metzad}, thereby under-specifying some of the variability in ecological time series due to adaptation. In so doing, classical AD ignores the potential phenomena of momentum and inertia during trait evolution, which has been well supported by evolutionary theorists \cite{simpson1944tempo}. 
The evolutionary field formulation represents a higher order approach to trait adaptation, which can describe much of this complexity in even the simplest predator-prey systems. It does this in purely adaptive terms through high-dimensional trait-based chaos, which can arise from even one to two traits. Our proposition therefore calls into question the orthodoxy of simple, low dimensional trait dynamics to adequately capture complexity (beyond purely periodic dynamics) and apparently random variation in ecological networks. 

The debate over the origin of variability in ecological time series has been ongoing since chaotic fluctuations were first observed in simple models of logistic population growth \cite{may1974biological}. Since then, a vast array of models have been produced in an attempt to characterise the most important elements of the erratic ecological time series and to ascertain the deterministic or stochastic nature of these complex signals \cite{schaffer1986chaos,berryman1989ecological,turchin1993chaos,hastings1993chaos}. Decades later debate still rages as to the very definition of chaos and noise in ecology \cite{dennis2003can,ellner2005can}, however much work has been done to suggest that explaining factors such as density dependence and persistent long-term autocorrelation will be necessary to produce a complete description of ecological dynamics \cite{berryman2001identifying,halley1996ecology}.

If higher order, high-dimensional deterministic dynamics are responsible for a portion of the apparent stochasticity seen in ecology, then such dynamics should share key characteristics with these stochastic processes. According to work by Halley and Inchausti \cite{inchausti2001investigating}, as much as 92\% of ecological time series exhibit spectral reddening or pink shift, the long-term pattern of increasing variation over time seen in ecological data. Coloured noise models have often been used to explain these typical patterns in ecological data \cite{halley1996ecology}. Stochastic noise processes and, in particular, coloured noise, share many important properties with chaotic dynamics, including finite correlation dimensions \cite{osborne1989finite} and even positive Lyapunov exponents in some cases \cite{dammig1993estimation}. Coloured noise is characterized in terms of its spectral properties; however, certain dynamical systems have frequency spectra which are qualitatively similar to noise \cite{manneville1980intermittency}. In addition, spectral reddening is seen as the hallmark of self-organised criticality in both physics \cite{bak1987self,bak1988self}, and more controversially in ecological models \cite{drake1999nature,sole1999criticality,fukami1999self}.

We show that selective field forces, acting at a distance in trait space, may be enough to superficially mimic many of these stochastic properties as well as attain a level of complexity comparable to real ecological data in the case of a simple predator-prey system. In Sec. \ref{just} we present a formal justification of the model framework and parameters, followed by (in Sec. \ref{chaos}) an exploration of the field model's dynamical and fractal chaotic behaviour in the chaotic, transient and aperiodic regimes. Here, we look specifically at intra-specific competition because of its established role in triggering instabilities and chaotic dynamics in population models \cite{may1974biological,may1976simple} and its importance in the variability of population data \cite{hanski1990density}. In Sec. \ref{fit}  we use historical field data on \textit{Oryctolagus cuniculus}, the European rabbit in Britain \cite{middleton1934periodic} and \textit{Lynx Canadensis}, the Canadian lynx \cite{elton1942ten} to determine whether the model fits with the qualitative behaviour of ecological systems. This was further investigated and tested in Sec. \ref{spec} using methods based on spectral analysis and the prominence of pink noise in ecological data with concluding remarks and further recommendations in Sec.\ref{disc}.
\section{Model Justification and Formulation}\label{just}
The model relies on density-dependence as the primary determinant of biological interaction frequency, both at the population and, by implication, at the evolutionary level. This mass action approach underpins classical and modern theories in physics (e.g. gravitational and solid state physics \cite{ross2013electrical}) and population ecology (e.g. Lotka-Volterra and AD models \cite{dercole2015ecology}). The evolutionary field model extends these ideas into evolutionary ecology by considering each biotic interaction as an exchange of fitness information between populations. An understanding of the model relies on interpreting individuals interacting within and between species, as mediators of an evolutionary force which is translated to adaptive change in a generalized trait space (an abstract representation of multiple independent, continuous traits). The model is also partially motivated by recognising the role played by density-dependence in both stochastic and dynamical complex behaviour in ecology (e.g. in inducing population level chaos in classical ecological models \cite{berryman1989ecological}). However, density considerations only describe the frequency, not the strength or type of individual interactions between members of two species. Both competitive and antagonistic interaction strengths have largely been measured in the past by trait matching \cite{kiester1984models}, in which distances between individuals' traits have some bearing on the strength of their interaction. This is motivated by the assumption that trait matching translates into stronger, more direct competition and more efficient consumption, or cooperation (in the case of mutualistic interactions). This does not mean that the two species are necessarily similar in phenotype as the traits relevant to each species in the interaction could differ in type or scale.

An appropriate measure must thus be chosen to quantify the degree of trait matching in a continuous trait space. The functional form of the measure is based on the concept of assortativity used in other adaptive dynamics models \cite{rossberg2010trophic,valdovinos2010consequences,zhang2013adaptive,Landi_et_al_13_SIAP}. In assortative models trait matching is measured as a Gaussian function of interaction strength based on the Euclidean norm ($||\cdot||$). Here, smaller Euclidean distance between traits implies stronger trait matching which decays exponentially with the square of the distance. From the inverse of this similarity measure we obtain a distance measure $d_{ij}$ from $i$ to $j$. The distance together with the direction vector $\mathbf{u_{ij}}$ informs the assumed topology of the trait space, which we will suppose for our purposes to be two dimensional, with trait vector $\mathbf{a_i}=(x_i,y_i)\in \mathbb{R}^2$ for species $i$; they are defined as:
\begin{align}
d_{ij}=e^{\frac{||\mathbf{a_i-a_j}||^2}{2}} \qquad \mathbf{u_{ij}}=\frac{\mathbf{a_j-a_i}}{||\mathbf{a_j-a_i}||}\label{distu}
\end{align}
We assume that the evolutionary or selective force experienced by a population in our model is dependent on this assortativity distance, which takes the form of an inverse Gaussian with standard mean and variance. Note that this is one form of the proposed matching distance and other forms may be applicable depending on context. 

The selective force itself can be derived from a field, $\mathbf{\Phi}$ mediated by interactions propagated by individuals within populations residing in a community of $N$ species. If we further suppose that these populations are mixed homogeneously then their interaction frequency could be assumed to be governed by mass action. We propose one possible way to decide on adaptive interaction strength is given by supposing that interaction strength deteriorates radially from the propagating population's position in trait space with distance defined by the assortitive distance (Eq. \eqref{distu}). If the same rules of point propagation apply as in physics then there exists an inverse square relationship between trait distance and adaptive interaction strength. Lastly, the nature and adaptive capacity (maximal strength) of interactions between species are specified by an $N\times N$ matrix $\mathbf{K}$ which is static in our formulation. From these proposed selection and frequency rules we arrive at a set of second order evolutionary field equations which determine the field strength $\mathbf{\Phi_i}=(\Phi^x_i,\Phi^y_i)$ experienced in trait space by species $i$ with population mass $m_i$.
\begin{align}
\mathbf{\Phi_i}=\sum\limits_{j=1}^{N}k_{ij}\frac{m_im_j}{d_{ij}^2}\mathbf{u_{ij}}\label{field}
\end{align}
The combination of the $k_{ij}$ and $k_{ji}$ interaction coefficients define the type (mutualistic, predatory, competitive etc.) and maximum potential interaction strength between species $i$ and $j$. These types are characterised by the effect that interaction with members of species $j$ usually has on the fitness and abundance of members of species $i$ and can be either antagonistic ($k_{ij}<0$) or beneficial ($k_{ij}>0$). Interactions resulting in changes at the population level translate into slower changes at the adaptive level through repulsive ($k_{ij}<0$) or attractive ($k_{ij}>0$) field effects on species $i$ with respect to $j$ in trait space. In general $k_{ii}<0$ and represents the negative intra-specific relationship between population density and the availability of environmental resources such as territory. 

These $k_{ij}$ factors determine the maximum potential interaction strength because of the bound imposed by the electrostatic or gravity-like inverse squared law given by the assortative distance in Eq. \eqref{adapt}. Since there are no constraints on the signs of the $k_{ij}$ and $k_{ji}$ pairs, this allows for the set up of pseudo-gravitational adaptive competitions (a combination of push pull and chase behaviours) between $i$ and $j$. These games operate similarly to systems of arbitrarily signed masses (population masses in our biological context) governed by electrostatic (or gravitational) attraction or repulsion \cite{bondi1957negative}.

The number of mutations which occur in a population shapes the capacity for evolutionary change to occur rapidly. This population mutation rate ($\theta_i$) is defined as the product of population mass ($m_i$) and individual mutation rate ($\mu_i$). We propose that under the force of selection, mutation defines the proportion of that force that can be converted into adaptive trait change, that is, in a mechanical sense, the population mutation rate $\theta_i$ can be likened to the quantity given by acceleration over force in mechanics ($\frac{a}{F}$ for force $F$ and acceleration $a$) or the inverse of mechanical inertia ($\frac{1}{m}$ for inertial mass $m$). This conception of evolutionary inertia is consistent with classical theories in AD including a special case of the canonical equation when one considers the second order time derivative \cite{dieckmann1996dynamical}.

However, overcoming inertia alone is not enough to lead to rapid adaptive change as there are many other constraints which slow the rate of adaptation in a population such as the time-dependent considerations of finite gene flow and generation time \cite{martin1993body,mooers1994metabolic}. These constraints may have a dampening effect on the speed of adaptive change by hampering the non-synonymous mutation rate \cite{ohta1993examination}. We summarize such effects as a frictional term which affects the rate of evolutionary change, choosing to model this evolutionary damping force by taking inspiration from models of fluid drag in physics, where friction scales with the square of the adaptive velocity (rate of trait change), with drag coefficient $f_i<0$. Completing the analogy with physics we suppose that adaptive change on a species $i$ is affected through the selective force on that species ($\Phi_i$) plus those frictional terms ($\mathbf{f}$) which resist against rapid adaptive change, scaled by the inverse of evolutionary inertia ($\theta_i$). The evolutionary equations of motion are thus given by Eq. \eqref{field}, to arrive at an equation for evolutionary acceleration in the traits $\mathbf{a_i}$,
\begin{align}
\frac{d^2\mathbf{a_i}}{d\tau^2}&=\theta_i\left[\mathbf{\Phi_i}+f_i\frac{d\mathbf{a_i}}{d\tau}\left|\frac{d\mathbf{a_i}}{d\tau}\right|\right]\label{adapt}
\end{align}
where adaptation operates at a slower time scale $\tau$ when compared to the community population dynamics.

The complete description of the eco-evolutionary system requires the specification of population dynamics equations which are influenced by changes in interaction strength brought about by adaptive change. Here we considered a two species, predator-prey system with prey mass $m_1$, and predator mass $m_2$. We again assumed homogeneous mixing of populations with simplified functional response (other more complex functional forms may be appropriate in other specific contexts). This leads to a simple Lotka-Volterra like set of population dynamics equations that are modulated by interaction strength, type and frequency in the same way as the field strength in Eq. \eqref{field}.
\begin{align}
\frac{dm_1}{dt}&=\left[r_1 + k_{11}m_1 + \frac{k_{12}}{d_{12}^2}m_2+ \frac{k_{1s}}{d_{1s}^2}m_s\right]m_1\label{k1}\\
\label{k2}\frac{dm_2}{dt}&=\left[r_2 + k_{22}m_2 + \frac{k_{21}}{d_{21}^2}m_1\right]m_2
\end{align}
(See Eq. \eqref{par1} and \eqref{par2} for parametrisation.) The two separate time scales $\tau$ and $t$ are such that $T=\frac{d\tau}{dt}<1$, but, since adaptation is assumed to occur according to a second order process, this means that $T^2$ is the time scale factor of relevance to the higher order dynamics. Here class $s$ represents a stationary resource which we introduced to maintain the community and can be considered a density-independent environmental resource or to be sufficiently abundant to be unaffected by prey consumption (i.e. it has a fixed density of $m_s=1$). It is also non-adaptive, existing at a fixed position at the origin in trait space (i.e. $\mathbf{a_s}=\mathbf{0}$). This model does not treat intraspecific competition, represented by $k_{ii}$ (defined in Eq. \eqref{k1} and \eqref{k2}), or death rate $r_i$ as adaptive. These parameters are thus independent of trait distance and matching in this particular model but could be made so in other more complicated versions of the model. 
\section{Dimensionality and Chaos}\label{chaos}
\begin{figure}[h!] 
	\centering
	\includegraphics[width=\linewidth]{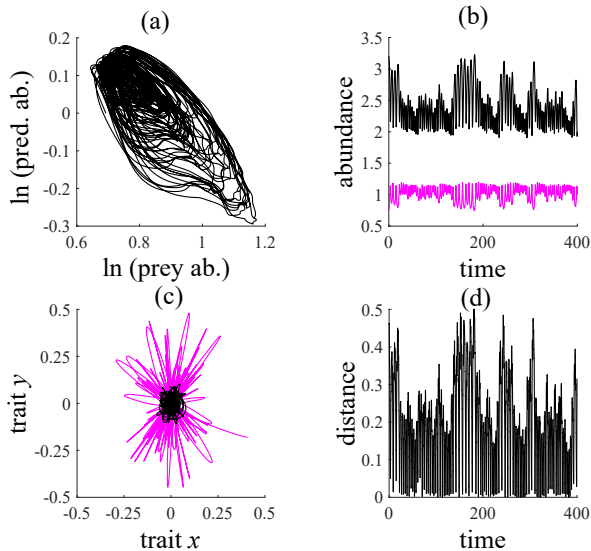}
	\caption{(Colour Online) Aperiodic predator-prey system behaviour at both the trait and population levels plotted for 400 time units (TU) in the case where prey intraspecific competition, $k_{11}=-0.5$. (a.) Log abundance-abundance and (b.) abundance-time for prey ($m_1$ - top, black curve) and predator ($m_2$ - bottom, magenta curve). (c.) Prey ($\mathbf{a_1}$ - black curve) and predator ($\mathbf{a_2}$ - lighter, magenta curve) trait space dynamics in a 2D trait space that shows aperiodic orbiting of the fitness optimum. (d.) Euclidean distance ($||\mathbf{a_1-a_2}||$) between predator and prey in trait space exhibiting stationary aperiodic behaviour.}
	\label{k1105}
\end{figure}

Numerical simulations of the two-species system, as defined in Sec. \ref{just} and parametrised in Eq. \ref{par1} and \ref{par2}, was carried out over a period of arbitrary time units, TU \cite{brown1989vode}. FIG. \ref{k1105} shows the results of simulation of a typical trajectory (after exclusion of $10^4$ TU of transient) for the case where the relationship between prey and predator intraspecific competition ($k_{ii}$) is $k_{11}=-0.5>k_{22}=-0.8$, with $1.1=\mu_1>\mu_2=1$. The system exhibits aperiodic cycling at the trait and population levels. Strong positive correlation ($\rho_{XY}=0.68$), evident between interspecies trait distance and prey abundance, indicates that an assortative force generated by a pseudo-gravitational field can lead to population fluctuations closely linked to co-evolutionary change. 

\begin{figure}[h!]
	\centering
	\includegraphics[width=\linewidth]{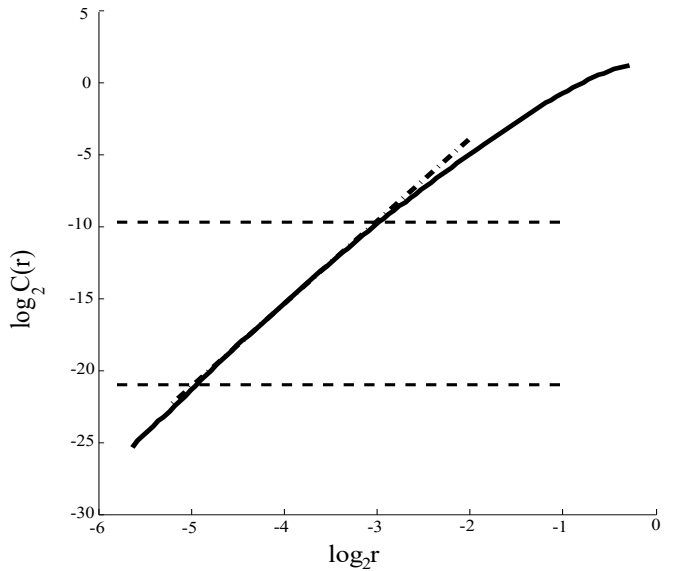}
	\caption{Log-Log plot of the correlation sum as a function of radius when prey density, $k_{11}=-0.5$. Horizontal, dashed lines indicate the bounds of the scaling region, where the log sum is near linear (in accordance with Eq. \eqref{dimlaw}) with slope $\hat{D_2}=5.7\pm 0.1$ (linear fit given by dashed line).}\label{fracfit}	
\end{figure}

The correlation dimension, $D_2$, is an established measure of the fractal dimension of chaotic attractors \cite{Grassberger1983} which is defined in terms of the distribution of randomly sampled points on the attractor. Calculating the correlation dimension, $D_2$, requires the computation of the correlation integral which can be approximated with real or simulated time series of size $N$ by the correlation sum $C(r)$. The correlation sum \cite{theiler1986spurious} is a weighted count of points from the series within a given radius $r$, of each other.
\begin{align*}
C(r)=\frac{2}{(N-c)(N-1-c)}\sum_{i=1}^{N}\sum_{j=1}^{i-c}H(r-||\mathbf{x_i-x_j}||)
\end{align*}
Here $H(x)$ is the Heaviside step function, $||\cdot||$ is the Euclidean norm and $\mathbf{x_i}$ are time-indexed points from a multidimensional time series. The integer $c$ defines a correlation length, and is used to exclude values that are close neighbours in time. The following relationship holds between the fractal dimension $D_2$, the radius $r$ and the correlation sum, $C(r)$:
\begin{align}
C(r)\propto r^{D_2}\label{dimlaw}
\end{align}
Due to this power law $D_2$ is approximated by the slope of the scaling region of the log-log plot of $C(r)$ versus $r$.  From FIG. \ref{fracfit} this gives an estimate of $D_2\approx5.7$ for the correlation dimension of the attractor. $D_2$ depends on the choice of scaling region and the value of parameters. For different values of the $k_{ij}$ interaction coefficients $D_2$ takes on a value $D_2\in[2.1,6.3]$ for prey intraspecific competition $k_{11}\in[-0.7,-0.4]$ and $D_2\approx 2.1$ for $k_{11}\in[-0.9,-0.8]$.

\begin{figure}[h!] 
	\centering
	\includegraphics[width=\linewidth]{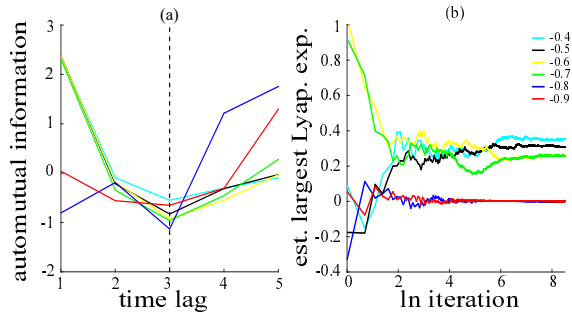}
	\caption{(Colour Online) (a.) Normalized, time-lagged mutual information (in bits, legend given in (b.)) for the prey population time series ($m_1$). The dotted line shows the first minimum (and thus the proposed delay time) as 3. (b.) Estimate of the Largest Lyapunov exponent ($\lambda_1$) estimated by the Wolf algorithm as the number of iterations (on a log scale) increases. The legend shows the parametrisations of prey intraspecific competition corresponding with the colour (tint) of the curve ($k_{11}$) between $-0.4$ and $-0.9$.}\label{lam}
\end{figure}

\begin{figure}[h!]
	\centering
	\includegraphics[width=\linewidth]{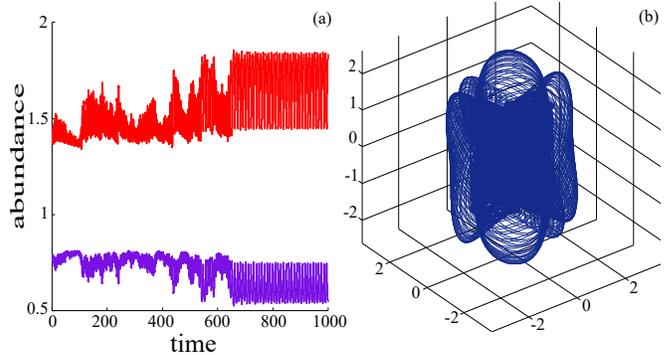}
	\caption{(Colour Online) (a.) Transient chaotic behaviour in species abundance when prey intraspecific competition, ($k_{11}$) is $-0.9$, showing prey ($m_1$ - top, red curve, coloured as in FIG. \ref{lam}) and predator ($m_2$ - bottom, magenta curve) time series. (b.) Time-delay reconstruction of a prey trait ($x_1$) time series (after transient), embedded in three dimensional space, exhibiting high-dimensional, non-chaotic (quasi-periodic) behaviour.}
	\label{trans}
\end{figure}

The presence of dynamical chaos in a time series can be detected using Wolf's algorithm \cite{wolf1985determining}. This algorithm uses the defining feature of chaos, exponential-time orbital divergence under small perturbations, to determine the largest Lyapunov exponent, $\lambda_1$. In order to show that dynamical chaos can be recovered from a variable more likely to be observed in the field, phase space reconstruction was carried out using prey abundance ($m_1$). The reconstruction was performed using the time-delay embedding theorem of Takens \cite{takens1981detecting}. An embedding dimension of 6 was chosen by taking the ceiling of the previous result for the correlation dimension and the time-delay was approximated using the first minimum of the auto-mutual information according to the method of Fraser and Swinney \cite{fraser1986independent} (FIG. \ref{lam}a). 
Investigation of $\lambda_1$ for a range of parametrisations of prey density-dependence, $k_{11}$, shows a change in behaviour for $k_{11}\le k_c\approx k_{22}=-0.8$ ($k_{22}$ is the coefficient of predator intraspecific competition) from a positive to negligible (possibly non-positive) $\lambda_1$ value, indicative of a bifurcation to chaos. This demonstrates a potential route to chaos for this predator-prey system dependent on the relationship between predator and prey density. The role of density-dependence in ecological chaos and in population stability and robustness has been widely supported in theory and simulation \cite{berryman1989ecological,may1974biological,may1976simple}. This behaviour may also represent an adaptive form of the Paradox of Enrichment presented in a highly controversial and influential paper by Rossenzweig \cite{rosenzweig1971paradox} in which de-stabilization of both populations can result from lowering the resource restrictions on prey.
Transient chaotic behaviour can result from the creation of an unstable chaotic manifold through the crisis (periodic) or the crisis-like (quasi-periodic) route to chaos \cite{grebogi1983crises}. Steady state dynamics following chaotic transients can be periodic, quasi-periodic or even include chaotic behaviour on a secondary attractor. Transient behaviour can be relatively persistent and can remain even after far exceeding the bifurcation value $k_c$ \cite{grebogi1985super}. In addition the steady state can be sensitive to any perturbations which could cause the system to re-enter a potentially long chaotic transient again \cite{faisst2003lifetimes}. 
The behaviour of the two species system for $k_{11}\ge k_c$ seems to exhibit such chaotic transient behaviour with a quasi-periodic steady state (FIG. \ref{trans}). The combination of even the lowest levels of noise and the presence of a chaotic repellor for $k_{11}>k_c$ could lead a simple predator-prey system to persist in a chaotic state.
\section{Model Fit to Prey Species Data}\label{fit}
\begin{figure}[h!]
	\centering
	\includegraphics[width=\linewidth]{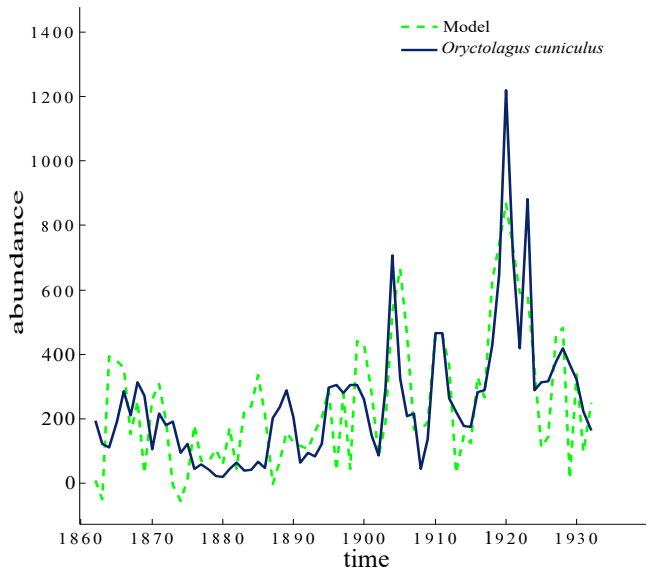}
	\caption{(Colour Online) Model fit (dotted, green curve, coloured as in FIG. \ref{lam}) to Middleton, \textit{Oryctolagus cuniculus}, data (solid) from Norfolk B, 1932-1862, using a fitted prey intraspecific competition coefficient ($k_{11}$) of $-0.7$ and a fitted sampling period of 7TU. These parameters were selected by windowed least squares fitting from simulated time series.}
	\label{rabbitfit}
\end{figure}

Ecological time series were obtained from work by Middleton on the European rabbit \textit{Oryctolagus cuniculus}, gathered annually in Norfolk (site B), Eastern lowland Britain, from 1862 to 1932 \cite{middleton1934periodic}. The European rabbit has been shown to dominate the diet of lowland red foxes (\textit{Vulpes vulpes}) in all seasons. The rabbit constitutes 74\% of mass ingested annually \cite{baker2006potential}. FIG. \ref{rabbitfit} shows the results of fitting the model using a loose minimum squared error (SE) approach, on $10^4$  data points (not including transient) of the simulated prey abundance. The time series were generated using the parametrisations already explored. The model was fitted to the data by sampling at different rates from the model (with a period of between 5TU and 70TU) using a moving window of the same size as the data set. The least-squares error was then calculated across all such windows and values of prey density dependence ($k_{11}$ from $-0.4$ to $-0.9$) to obtain the best fit for the data.
\section{Noise Generation and Spectral Comparison of Model with Data}\label{spec}
Pink noise describes a family of random signals termed coloured noise which contaminate a vast range of real world signals, including the majority of ecological time series data \cite{pimm1988variability,rohani2004colour,halley2004increasing}. The presence of pink noise is characterised by long-term correlations in the series. It is defined by the inverse relationship between the frequencies present in the underlying signal $x(t)$ of the time series and the amount of energy (variation) present at each frequency, known as the power spectral density, (PSD) of the signal at $f$, denoted $S_{xx}(f)$. The relationship can be expressed as
\begin{align}
S_{xx}(f)\propto \frac{1}{f^\alpha}\label{npl}
\end{align}
where $0\le\alpha<3$ is the noise scaling exponent which determines the rate at which power drops off with frequency. Typically, any signal for which $2>\alpha>0$ is said to be reddened, while noise where $\alpha\approx 1$ is known as pink noise.  In contrast, white noise is a random signal with a constant power level over all frequencies i.e. $S_{xx}(f)\propto 1$. 
A property that distinguishes pink from white noise and which makes pink noise even more interesting from a dynamical perspective is that pink noise possesses finite fractal dimension dependent on the value of $\alpha$. This makes it more difficult to distinguish from the underlying dynamics with the use of fractal analysis \cite{osborne1989finite}. This relationship is restricted to $1<\alpha<3$.
\begin{align*}
D_2(\alpha)=\frac{2}{\alpha-1}
\end{align*}
A method for estimation of the $\alpha$ noise exponent, $\hat{\alpha}$, in short, stationary ecological time series was followed, as presented by Miramontes and Rohani \cite{miramontes2002estimating}. This method has been used effectively to identify exponents in series as short as 40 data points \cite{rohani2004colour}. The standard method utilizes the direct estimation of PSD via the absolute square of the Fourier Series. However, a more accurate method for PSD estimation was used here, the so-called Multitaper approach proposed by Thomson \cite{thomson1982spectrum}. The Multitaper method is a non-parametric method of PSD estimation which reconstructs the spectrum by averaging over pairwise-orthogonal windowed segments of the original series (which are thus statistically independent). This method has a number of advantages over the direct Fourier transform in that it is not dominated by bias, and the averaging of orthogonal data windows has the effect of smoothing out some of the noise caused by sample size limitations.
The value of $\hat{\alpha}$ for the Middleton data was calculated as $0.948$ with a 95\% confidence interval of $CI=[0.383,1.51]$. In comparison, for the fitted data $\hat{\alpha}=1.13\in CI$. This suggests agreement at the spectral level, not just between the model and data, but also with previous results for long ecological series. 
Spectral similarities persist for longer, chaotic time series as well. 

\begin{figure}[h!]
	\centering
	\includegraphics[width=\linewidth]{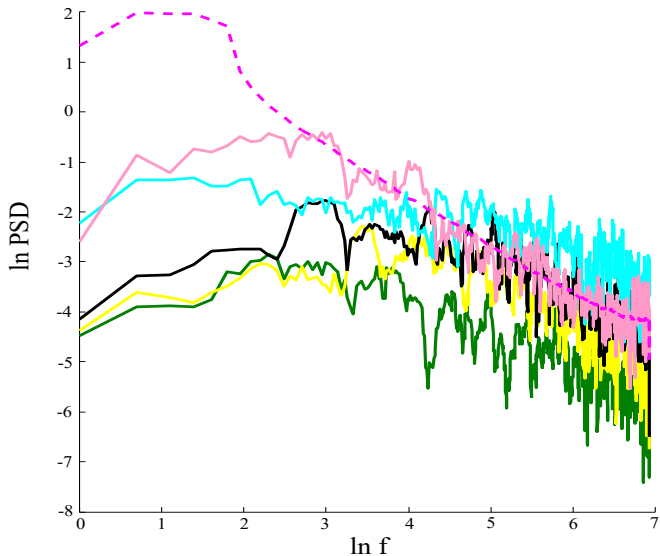}
	\caption{(Colour Online) Log-log spectral density (variance in each frequency) reconstructed from simulated prey population series ($m_1$ - colour as in FIG. \ref{lam}) at a sampling period of 7/TU (totalling 1429 data points per parametrisation), determined by estimation from Middleton data. Includes spectra from pink noise that has been smoothed (dotted, magenta curve), by averaging 1000 individual signals, as well as a single representative realization in solid pink (top-most, solid curve), both with the same total variance (power) as the model time series with intraspecific competition ($k_{11}$) equal to $-0.5$. Superficially similar decreasing linear trend between pink and chaotic spectra indicates variational similarities.}
	\label{logspectra}
\end{figure}

Simulated pink noise data was generated using the digital signal generation method produced by Kasdin, and also independently by Hoskings \cite{kasdin1995discrete,hosking1981fractional}, which relies on convolution of a white noise series with a transfer function. In FIG. \ref{logspectra}, negative linear trend dominates in the log-log PSD plots for all chaotic intraspecific competition ($k_{11}$) parametrisations.  The chaotic signals mimic the simulated pink noise data in their qualitative behaviour (with comparable trend and slope) at all but the lowest frequencies where higher variation is present in the pink series, however, this discrepancy may become difficult to notice in short ecological series. The shared negative linear trend and slope in the higher frequency log spectrum indicates a shared power law variance drop-off relationship (see Eq. \eqref{npl}) between the chaotic model and noise signals.

A non-parametric method to distinguish chaos from a series generated by any coloured noise process has been proposed by Kennel and Isabel, which uses a Kolmogrov-Smirnov statistic derived from simulating the prediction error of a large number of surrogate data. The surrogate data are based on the original query series with Gaussian noise added in the frequency domain \cite{kennel1992method}. The Kolmogrov-Smirnov statistic should behave as a standard normal random variable under the null hypothesis of no difference in generating distribution. Using a prediction step size of one, this method fails to distinguish our fitted model time series from coloured noise using a two-sided z-test ($z=-0.084>z_{0.05}$). This is in comparison with a value of $z=-0.307>z_{0.05}$ for the \textit{Oryctolagus cuniculus} field data. 
Importantly, it can be shown that certain ecological time series of predator-prey systems have compositions significantly different from noise at the 99\% confidence level, suggesting that other processes (e.g. periodic or quasiperiodic dynamics) dominate the spectrum. Data obtained from historical fur sales records of the Canadian lynx, \textit{Lynx canadensis}, in the MacKenzie River area of Canada, were obtained for the years 1821 to 1934 \cite{elton1942ten}. The test statistic calculated for these data was $z=-2.86<z_{0.005}$. This result means that, despite data size limitations, the test has sufficient power to detect significant deviations from coloured noise in some cases. Interestingly, fitting of the Canadian lynx data by the same process as used for the Middleton data, gives a non-chaotic parametrisation of best fit with intaspecies competition, $k_{11}=-0.8$ (in the quasiperiodic region). This demonstrates the potential of adaptive models to explain different kinds of variation in data.
\section{Discussion}\label{disc}
We have presented an eco-evolutionary model inspired by field ideas in physics which, using sufficiently fast evolving traits (such as behavioural or other phenotypically plastic adaptations to predation), can explain some of the complex patterns of population variability seen in simple ecological systems. As one of a large class of similar models our model is able to match the qualitative behaviour of specific ecological time series (\textit{Oryctolagus cuniculus}). However, what sets this model apart is that it demonstrates an instance in which high-dimensional adaptive models can have variational distributions characteristic of ecological systems in both the specific and abstract cases. The characteristic red shift seen in the spectral composition of our model is consistent with prior results for the majority of ecological time series and moreover shows that such properties need not necessarily arise from purely stochastic processes in ecology. Our findings do not supersede stochastic explanations but do show how high-dimensional, deterministic ecological dynamics (based on second order adaptive dynamics) and environmental stochasticity could, under certain conditions, sustain and reinforce each other leading to well-recognised patterns of complex variation found in ecology. 

The role played by intraspecific competition in triggering a bifurcation to dynamical chaos is notable in that it is in agreement with previous theoretical and observational research on the relationship between stability, variability and density dependence \cite{berryman1989ecological,may1974biological,may1976simple}. It also shows an alternative route to an effect similar to the paradox of enrichment \cite{rosenzweig1971paradox}, since decreased strain on the prey population leads to population instability in the prey and the system as a whole.

The combination of an appropriate trait matching and interaction frequency measure are the key components of an evolutionary field model. The model and functional forms we have chosen here were based on previous assumptions and theories in AD and population ecology and many other potentially viable field models of the same general form as Eq. \eqref{field} may exist which might not exhibit the same spectral properties. However, the versatility of the framework means that these models might also be made more application specific, based on the particular selective and demographic factors appropriate to the ecosystem of interest. However, since the form of the model is very general it is possible that not all formulations may behave in a biologically realistic way, implying a need for functions to be carefully chosen to fit the biological scenario.

The evolutionary field framework represents a possible explanation for commonly observed perturbations near to fitness optima, however, it is a significant departure from more orthodox AD theories. The most striking difference is a lack of an explicit selection gradient. The selection gradient is defined as the local fitness slope which populations experience due to the difference in their adaptive traits (taken to be the mean phenotype in a relatively genotypically homogeneous population) relative to all other populations in the community  \cite{dieckmann1996dynamical,deriad}. Although effects similar to the selection gradient might be explained by topological deformations of trait space caused by an evolutionary field, many of the emergent phenomena in AD, such as evolutionary branching \cite{geritz1998evolutionarily,geritz1997}, have yet to be described fully in this context \cite{br_bif_paper,dercole2016transition}.

Despite these difference in approach, the principle that ecological models demand greater capacity for complexity than has currently been achieved is evident and remains a major challenge for AD to overcome. The potential for field thinking in ecology may represent an underlying mechanical symmetry between ecology and physics and provide a new conceptual source for classical, game-theoretic models. Such models could be more easily extended to higher dimensional systems including those with multiple, species and traits, with less fine tuning than is generally required from current adaptive dynamics approaches.
\begin{acknowledgments}
This research was first presented as part of a BSc (Hons) thesis at the University of Stellenbosch with helpful advice from members of Stellenbosch University's Mathematics Department, notably Farai Nyabadza, the African Institute for Mathematical Sciences (AIMS), notably Jeff Sanders and Zoe Wyatt and the South African Centre for Epidemiological Modelling and Analysis (SACEMA), notably Brian Williams. The research was funded by the National Research Foundation (NRF) of South Africa through the South African Research Chair Initiative and the Competitive Program for Rated Researchers (NRF grants 81825 and 76912) with additional backing from SACEMA and AIMS.
\end{acknowledgments}
\appendix*
\section{Parametrization}
The full model parametrisation is presented here in a matrix format similar to the Lotka-Volterra models on which the population component is based.
\begin{align}
\mathbf{K}=\;\;\begin{blockarray}{ccc c}
1 & 2 & s \\
\begin{block}{(ccc)c}
\;\star &\; -1 &\; 2 \;\;& \;1 \\
\;0.5 & \;-0.8 &\; 0 \;\;& \;2 \\
\;0 & \;0 & \;0 \;\;& s \\
\end{block}
\end{blockarray}\qquad \mathbf{r}=\;\;\begin{blockarray}{c c}
 &  & \\
\begin{block}{(c)c}
\;-0.002 & \;\;1 \\
\;-0.003 & \;\;2 \\
\end{block}
\end{blockarray}\label{par1}
\\ \mathbf{\mu}=\;\;\begin{blockarray}{c c}
&  & \\
\begin{block}{(c)c}
\;1.1 & \;\;1 \\
\;1 & \;\;2 \\
\end{block}
\end{blockarray}\qquad T=0.5 \qquad \mathbf{f}=\mathbf{1}\label{par2}
\end{align}
The $\star$ for $k_{11}$ is a place holder that denotes the changing value of prey density-dependence ($k_{11}$) between sections and figures. Curves (such as spectral and prey density, $m_1$, plots) derived from simulations where $k_{11}$ takes on a specific value, have a corresponding colour scheme. This colour scheme for $k_{11}$ is illustrated in Table \ref{tab} with values repeated in the order they appear (names of colours are tinted according to their grey-scale value). Other colours are specified when considering other system (or noise model) variables for a specific value of the density-dependence ($k_{11}$).
\begin{table}[htbp]
	\centering
	\caption{Colour scheme for prey density dependence ($k_{11}$)}
	\begin{tabular}{rrrrrrrr}
		\toprule
		$k_{11}$ \vline& $-0.5$   & $-0.4$   & $-0.5$   & $-0.6$   & $-0.7$   & $-0.8$   & $-0.9$ \\
		\colrule
		colour \vline& black & \textcolor{gray-cyan}{cyan}  & black & \textcolor{gray-yellow}{yellow} & \textcolor{gray-green}{green} & \textcolor{gray-blue}{blue}  & \textcolor{gray-red}{red} \\
		\botrule
	\end{tabular}%
	\label{tab}%
\end{table}%
\bibliography{cites}

\end{document}